\documentclass[epj]{svjour}
\usepackage{graphicx}
\usepackage{amssymb}
\usepackage[latin1]{inputenc} 
\usepackage{url}

\begin{document}




\title{Exclusive measurement of coherent $\eta$ photoproduction from the
deuteron}


\author{J. Weiß\inst{1} \and
	P. Achenbach\inst{2} \and
	J. Ahrens\inst{2} \and
	J.R.M. Annand\inst{3} \and
	R. Beck\inst{2} \and
	V. Hejny\inst{4} \and
	J.D. Kellie\inst{3} \and
	V. Kleber\inst{4} \and
	M. Kotulla\inst{1} \and
 	B. Krusche\inst{5} \and
	V. Kuhr\inst{6} \and
	R. Leukel\inst{2} \and
	V. Metag\inst{1} \and
	V. M.\ Olmos de Léon\inst{2} \and
	F. Rambo\inst{6} \and
	A. Schmidt\inst{2} \and
	U. Siodlaczek\inst{7} \and
	H. Ströher\inst{4} \and
	F. Wissmann\inst{6} \and
	M. Wolf\inst{1}}

\institute{II.\ Physikalisches Institut,
Justus-Liebig-Universität Gießen, Heinrich-Buff-Ring 16, D-35392
Gießen
\and
Institut für Kernphysik,
Johannes-Gutenberg-Universität Mainz, Johannes-Joachim-Becher-Weg 45,
D-55099 Mainz
\and
Department of Physics and Astronomy, University of Glasgow, Glasgow
G128QQ, UK
\and
Institut für Kernphysik, Forschungszentrum Jülich GmbH,
D-52425 Jülich
\and
Departement für Physik und Astronomie, Universität  Basel,
Klingelbergstrasse 82, CH-4056 Basel
\and
II.\ Physikalisches Institut,
Georg-August-Universität Göttingen, Bunsenstr.\ 7-9, D-37073
Göttingen
\and
Physikalisches Institut,
Eberhard-Karls-Universität Tübingen, Auf der Morgenstelle 14, D-72076
Tübingen}

\mail{Bernd.Krusche@unibas.ch}

\date{\today}

\abstract{%
Coherent photoproduction of $\eta$ mesons from the deuteron
has been measured from threshold up to $E_\gamma\approx$ 750~MeV using the
photon spectrometer TAPS at the tagged photon facility at the Mainz
microtron MAMI. For the first time, differential coherent cross
sections have been deduced from the coincident detection of the $\eta$ meson 
and the recoil deuteron. A missing energy analysis was used for the 
suppression of background events so that a very clean identification of
coherent $\eta$-photoproduction was achieved.
The resulting cross sections agree with previous experimental results except
for angles around 90$^o$ in the $\gamma d$ cm system where they are smaller. 
They are compared to various model calculations.}


\maketitle


\section{Introduction}
\label{sec:intro}

The structure of the nucleon and its excited states is one of the central
issues of nonperturbative Quantum Chromodynamics. Meson photoproduction has 
emerged as an excellent tool for experimental investigations providing 
detailed resonance properties which are the ideal testing ground for modern
hadron models. One of the major difficulties for the experimental study of
excited nucleon states is that they are closely spaced and, due to their 
hadronic decay channels, have large widths which result in a significant 
overlap. Furthermore non-resonant background contributions like nucleon Born 
terms or vector meson exchange complicate the interpretation of photoproduction
reactions. However, there are a few cases where, due to the different couplings 
of the resonances to the initial photon - nucleon state and the final nucleon -
meson states, certain resonances strongly dominate photoproduction reactions.   
The best known example is the dominance of the low lying P$_{33}(1232)$
$\Delta$-resonance in $\pi^o$-photoproduction. In the so-called second resonance
region which comprises the P$_{11}$(1440), D$_{13}$(1520) and S$_{11}$(1535)
resonances the largest contribution to pion photoproduction comes from the
excitation of the D$_{13}$ due to the large photon coupling. On the other hand 
$\eta$-photoproduction in this energy range proceeds almost exclusively via the
excitation of the S$_{11}$ resonance \cite{Krusche_1,Krusche_2}. The decay
branching ratio of this resonance into $N\eta$ is approximately 50\% \cite{PDG}
while for the D$_{13}$ it is less than 1\% \cite{Tiator}. Eta photoproduction 
is therefore the best reaction for the study of the S$_{11}$ resonance on the 
free nucleon \cite{Krusche_1} and in nuclear matter \cite{Roebig,Yorita}.
The negligible contribution of the P$_{11}$- and D$_{13}$-resonances to
$\eta$-photoproduction is easily understood since these decays must proceed 
via $\eta$ - nucleon pairs with relative orbital angular momenta of $l=1,2$ 
which are strongly suppressed close to threshold.  However, it is not 
understood why the contribution of the S$_{11}(1650)$ is also very small.       
   
Eta photoproduction from the proton has been used to determine the basic 
properties of the S$_{11}$(1535) resonance, such as its mass, width and in 
particular the electromagnetic helicity amplitude $A^{p}_{1/2}$ 
\cite{Krusche_1,Krusche_2}. The isospin structure of the electromagnetic 
excitation of the S$_{11}(1535)$ resonance was investigated in experiments 
on $\eta$ photoproduction from the deuteron \cite{Krusche_3,Hoffmann}. 
Quasifree $\eta$-photoproduction from the deuteron can be used to deduce the 
cross section of the $n(\gamma ,\eta)n$ reaction. However the complete isospin 
decomposition requires in addition the measurement of coherent photoproduction 
from an isospin I=0 nucleus. Since the excitation of the S$_{11}$-resonance 
via the $E_{0+}$ multipole involves a spin-flip transition, the I=0, J=1 
deuteron is the ideal target nucleus for this purpose. The isospin composition 
can be deduced from these measurements via:
\begin{eqnarray}
  \label{eq:isospin}
  \sigma_p & \sim & |A^s_{1/2} + A^v_{1/2}|^2 = |A^p_{1/2}|^2 \\
  \sigma_n & \sim & |A^s_{1/2} - A^v_{1/2}|^2 = |A^n_{1/2}|^2 \\
  \sigma_d^\mathrm{coh} & \sim & |A^s_{1/2}|^2
\end{eqnarray}
where $A^s_{1/2}$ denotes the isoscalar and $A^v_{1/2}$ the isovector 
amplitude. 

While quark models and results from pion photoproduction predicted a dominant 
isovector component of the S$_{11}(1535)$ excitation, an early 
experiment found a large coherent eta production cross section from the  
deuteron \cite{Anderson}, requiring a large isoscalar contribution. More 
recent experiments found the cross section to be much lower 
\cite{Krusche_3,Hoffmann} and extracted $A^s_{1/2} < A^v_{1/2}$, as quark 
models have predicted. Furthermore, from inclusive and exclusive measurements 
of quasi-free $\eta$ photoproduction from the deuteron and $^4$He, 
a cross section ratio of 
$|A_{1/2}^n|^2/|A_{1/2}^p|^2=\sigma_n/\sigma_p \approx 0.66$ 
has been deduced \cite{Krusche_3,Hoffmann,Hejny}. Taking into account the 
small coherent cross section from the deuteron a ratio of 
$|A^s_{1/2}|/|A^p_{1/2}|=0.09$ was extracted in \cite{Krusche_3,Hoffmann}.

On the other hand models reproducing the measured cross section of coherent
$\eta$ photoproduction from the deuteron \cite{Fix,Kamalov} require
$|A^s_{1/2}|/|A^p_{1/2}| = 0.22-0.26$, a value in conflict with the
experimental result above (cf. sec.\ \ref{results}). This discrepancy
motivated further studies of coherent photoproduction of $\eta$ mesons
from the deuteron. Ritz et al. \cite{Ritz} recently suggested that the problem
is due to contributions from hadronic rescattering which give rise to
a complex, energy-dependent phase relation between the extracted amplitudes. 
They deduce from a fit of their model to the data 
$A_{1/2}^p=(120.0-i66.1)\times10^{-3}GeV^{-1/2}$,
$A_{1/2}^n=(-114.0-i1.7)\times10^{-3}GeV^{-1/2}$
which are consistent with
$|A_{1/2}^n|^2/|A_{1/2}^p|^2\approx 0.66$ {\it and} 
$|A_{1/2}^s|/|A_{1/2}^p|\approx 0.25$.

\section{Experimental setup and analysis methods}
\label{experiment}

Coherent photoproduction of $\eta$ mesons has been measured at the Glasgow
tagged photon facility \cite{Anthony} at the Mainz microtron MAMI 
\cite{Walcher,Ahrens} using the photon spectrometer TAPS. A 
quasi-monochromatic photon beam of energies up to 818~MeV was produced via 
bremsstrahlung tagging. The TAPS detector \cite{Novotny,Gabler} consisted of 
6 blocks each with 64 hexagonal cross section BaF$_2$ crystals and a 
forward detector with 120 BaF$_2$ crystals. The 6 blocks were located 
in a horizontal plane around the target at angles of $\pm50^\circ$, 
$\pm100^\circ$, and $\pm150^\circ$ with respect to the beam axis. The forward 
detector covered a polar angular range of $5^\circ<\theta<20^\circ$. This
setup subtended $\approx$ 30\% of the full solid angle. All BaF$_2$ 
modules were equipped with 5~mm thick plastic detectors for the identification 
of charged particles. The target was a 10~cm long, 4 cm diameter, liquid
deuterium filled cell. A detailed description of the setup is given in 
\cite{Hejny}.

\begin{figure}[h]
\resizebox{0.5\textwidth}{!}{
\includegraphics{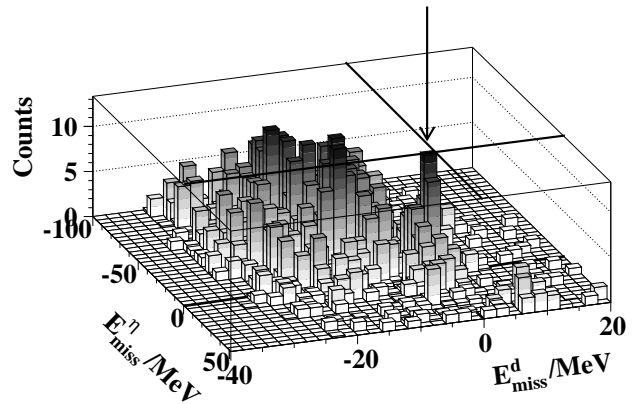}
}
\caption{Missing energy of both detected particles calculated
according to eq. \ref{eq:cohemissd} and \ref{eq:cohemisseta}. The peak
corresponding to coherent events is marked with an arrow.}
\label{fig:emiss}
\end{figure}

Eta mesons were detected via their two photon decay channel and identified in 
a standard invariant mass analysis using the measured photon energies and 
angles as input. An invariant mass resolution of $\approx$ 60~MeV (FWHM) was 
achieved for the $\eta$ signal. The photons were identified by time of flight 
and pulse shape analysis, while deuteron identification relied on energy and 
energy loss information from the plastic and BaF$_{2}$ scintillators 
(see \cite{Hejny} for details).  A missing energy analysis which utilises the 
kinematical overdetermination of the reaction was performed to obtain clear 
identification of coherent events. The difference of the c.m. energies of both 
the $\eta$ ($E^\ast_\eta$) and the deuteron ($E^\ast_d$) was calculated for 
this purpose from initial and final state kinematic quantities as follows:

\begin{eqnarray}
  E_\mathrm{miss}^d & = & E^\ast_d(E_\gamma, m_d) - E^\ast_d(E_d^\mathrm{lab},
  \mathbf{p}_d^\mathrm{lab}) 
  \label{eq:cohemissd}\\
  E_\mathrm{miss}^\eta & = & E^\ast_\eta(E_\gamma, m_d) - 
  E^\ast_\eta(E_\eta^\mathrm{lab},
  \mathbf{p_\eta^\mathrm{lab}}) 
  \label{eq:cohemisseta}
\end{eqnarray}

where $E_\gamma$ denotes the incident photon beam energy, $m_d$ the deuteron 
mass, $E_{d,\eta}^\mathrm{lab}$ and $\mathbf{p}_{d,\eta}^\mathrm{lab}$ the 
measured energies and momenta of deuteron and $\eta$ meson. The resulting 
two-dimensional missing energy distribution (see fig. \ref{fig:emiss}) allows 
a very efficient discrimination of the signal from remaining background. The 
peak at zero (marked with an arrow), which is clearly separated from the
background, corresponds to the coherent events.

The acceptance of TAPS for $\eta$-meson decays into two photons in the energy 
range of interest covers the full polar angle of the $\eta$-mesons. 
\begin{figure}[h]
\resizebox{0.5\textwidth}{!}{
\includegraphics{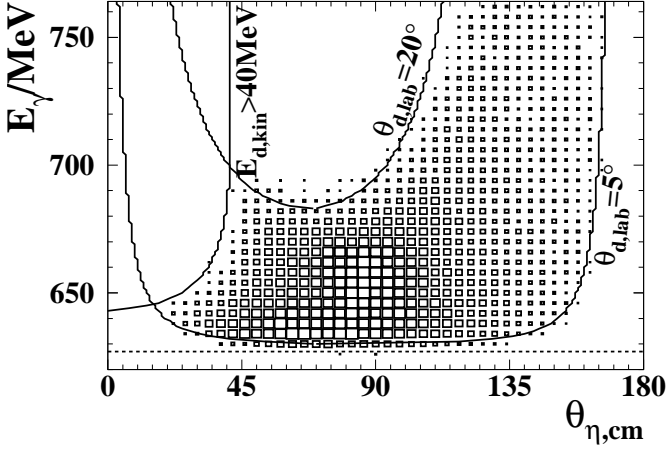}
}
\caption{Simulated acceptance for coherent $\eta$-production in TAPS. 
The boundaries correspond to the borders of the forward detector and the 
minimum energy required for the deuteron to be detected. The efficiency ranges
from 1.5 to 2.8\%.}
\label{fig:accept}
\end{figure}
However, for the deuteron there are regions, where the particle escapes
detection. Fig. \ref{fig:accept} shows the acceptance of TAPS for coherent 
events as a function of photon beam energy and $\eta$ polar angle in the c.m. 
system. The boundaries are determined by the angular range covered by the 
forward detector ($5^\circ<\theta<20^\circ$) and the minimum kinetic energy 
required for the detection of the deuterons ($E_{kin}>40$MeV). 

\section{Results}
\label{results}
\begin{figure}[h]
\begin{center}
\resizebox{0.40\textwidth}{!}{
\includegraphics{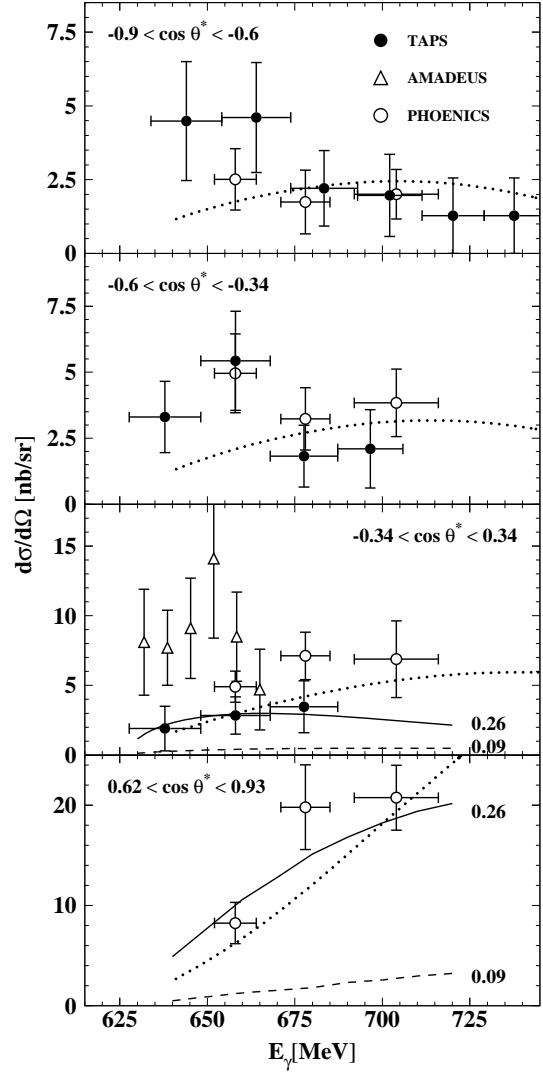}
}
\end{center}
\caption{Excitation function for coherent photoproduction from the deuteron
measured with TAPS
({\Large$\bullet$} The horizontal error bars indicate the width of the
incident photon energy bins) for different angular regions in
comparison with experimental data from PHOENICS ({\Large$\circ$}) and
AMADEUS ($\bigtriangleup$) \cite{Hoffmann} The PHOENICS data for backward
angles correspond to slightly different angular bins then the TAPS data
(-1,-06; -0.6,-0.2). The lowest plot shows the forward region, where TAPS has 
no acceptance. Calculations of Fix and Arenh\"ovel \cite{Fix} for various 
values of $|A_{1/2}^s|/|A_{1/2}^p|$ (solid: 0.26, dashed: 0.09) and Ritz 
and Arenh\"ovel \cite{Ritz} (dotted) are compared to the data.}
\label{fig:anr}
\end{figure}

\begin{figure}[t]
\begin{center}
\resizebox{0.40\textwidth}{!}{
\includegraphics{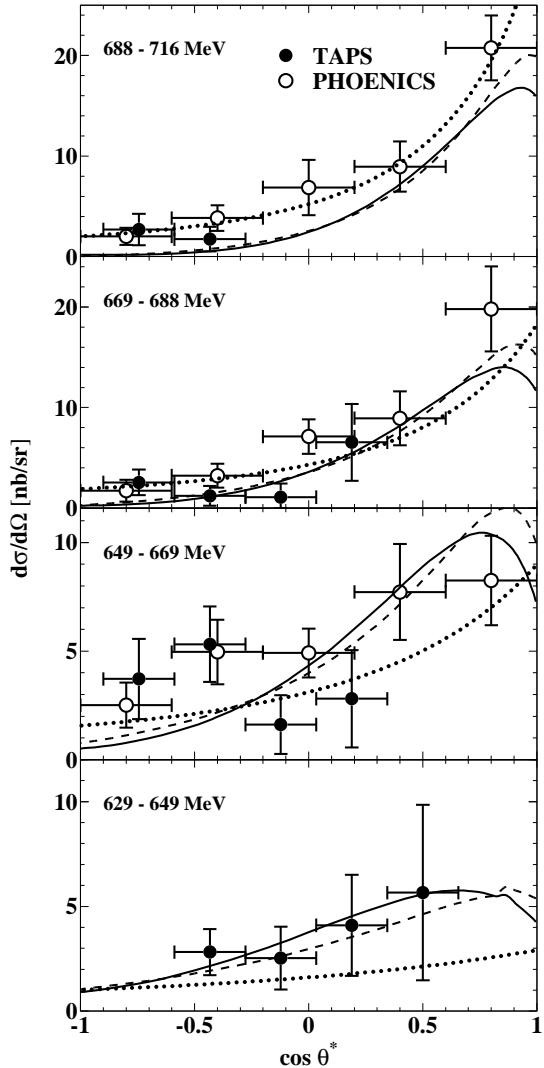}
}
\end{center}
\caption{Angular distributions of coherent $\eta$ mesons produced from
the deuteron for different photon beam energy regions ({\Large
$\bullet$} TAPS, {\Large$\circ$} PHOENIX, horizontal error bars
represent the angular bin width). Calculations are PWIA (dashed) and
DWIA (solid) from Kamalov et al. \cite{Kamalov} for
$|A_{1/2}^s|/|A_{1/2}^p|=0.22$ and from Ritz et al. \cite{Ritz} (dotted).}
\label{fig:wdiff}
\end{figure}

The measured differential cross sections for coherent $\eta$ photoproduction
from the deuteron are summarized in figs. \ref{fig:anr} and \ref{fig:wdiff}. 
The results of measurements performed at ELSA \cite{Hoffmann} with the
AMADEUS and PHOENICS detectors are shown for comparison. 
The identification of coherent $\eta$-production in all previous experiments 
relied completely on a missing mass analysis of the detected deuterons. 
The $\eta$-mesons were not identified via their invariant mass which increased 
counting statistics through the inclusion of the $\eta\rightarrow 3\pi$ decay 
channels. By contrast, the TAPS experiment detected deuterons {\it and} 
$\eta$-mesons in coincidence. This improves substantially the suppression of 
background, in particular from double $\pi^0$ production.

The resulting differential cross section for the backward angle bins shows a 
rise at production threshold, falling off again above 675 MeV photon energy.  
The comparison of the differential cross sections around 
$\theta_\eta^\ast=90^\circ$ measured by the different experiments shows, that 
the TAPS result is about 50\% lower than that of PHOENIX (fig. \ref{fig:anr}). 
However, both results could be consistent within the statistical uncertainties. 
The difference to the AMADEUS data near threshold is much larger. The lowest 
part of fig. \ref{fig:anr} shows the angular range of 
$0.62 < \cos\theta_\eta^\ast < 0.93$, which is not covered by the TAPS 
acceptance. Fix and Arenh\"ovel \cite{Fix} have compared their calculations 
with the PHOENICS data in this angular region. Using $|A_s|/|A_s+A_v|= 0.09$ 
(cf. sec. \ref{sec:intro}), the magnitude of the measured cross section is not 
reproduced. Instead a ratio of 0.26 is needed to fit the data. In the region 
around $\theta_\eta^\ast=90^\circ$, the model describes the TAPS data only 
with the same large value of the amplitude ratio.  The calculation of Ritz 
et al. \cite{Ritz} agrees with the TAPS and PHOENICS data quite reasonably 
except for the lowest incident photon energies at backward angles.

The situation is similar for the $\eta$ angular distributions. Fig. 
\ref{fig:wdiff} shows the coherent $\eta$ angular distributions in the photon 
deuteron c.m. system for different ranges of the incident photon energy. 
The data are compared to a calculation of Kamalov et al. \cite{Kamalov}. 
This model uses the coupled channel method in plane wave (PWIA) and distorted 
wave (DWIA) impulse approximations. The model predictions agree with the data 
within the statistical uncertainty assuming $|A_s|/|A_s+A_v|= 0.22$. With the 
exception of the lowest inicdent photon energy the calculation of Ritz et al. 
\cite{Ritz} is in similar agreement with the data.

\section{Conclusion}
\label{conclusion}

Coherent $\eta$ photoproduction from the deuteron has been measured by 
coincident registration of the recoil deuteron {\it and} the $\eta$ meson 
which very efficiently suppresses possible background contributions. 
The cross sections agree for most kinematical regimes with the results
from a previous experiment \cite{Hoffmann} but they are smaller in the angular
range around 90$^o$ in particular when compared to the AMADEUS data. 
Calculations in the framework of different models are in reasonable agreement 
with the data when the contribution of the isoscalar amplitude is choosen in 
the range $|A_{1/2}^s|/|A_{1/2}^s+A_{1/2}^v|\approx 0.22-0.25$. This relatively
large  isoscalar contribution can be reconciled with the measured 
proton/neutron cross section ratio 
$|A_{1/2}^n|^2/|A_{1/2}^p|^2=\sigma_n/\sigma_p \approx 0.66$
if rescattering contributions give rise to a large relative phase between
the proton and neutron amplitudes \cite{Ritz}. However, the interpretation of 
the $d(\gamma ,\eta)d$ reaction involves the isospin composition of the 
S$_{11}$ excitation and at the same time the correct treatment of nuclear 
effects such as meson rescattering contributions. The best way to disentangle 
these aspects is the investigation of the coherent process for nuclei with 
different quantum numbers which act as spin/isospin filters. Particularly 
interesting is a comparison of the I=0, J=1 deuteron to the I=J=1/2 nucleus 
$^3$He. In the latter case, as in the process on the free nucleon, the reaction
is dominated by the large isovector, spin-flip amplitude. Consequently, 
relatively large cross sections are expected and the reaction is ideally 
suited to test the model treatment of the FSI effects. This reaction has been 
measured in a very recent experiment with TAPS at MAMI and the data are 
currently under analysis.

\section{Acknowledgements}

The authors gratefully acknowledge the outstanding support of the
accelerator group of the Mainz microtron MAMI, as well as the
other technicians and scientists of the Institut für Kernphysik at the
Universität Mainz. We would like to thank H. Arenh\"ovel, A. Fix, S. Kamalov
and F. Ritz for discussions and correspondence.

This work was supported by Deutsche
Forschungsgemeinschaft (SFB 201) and the U.K.\ Engeneering and
Physical Sciences Research Council.







\end{document}